\documentclass[11pt,twocolumn,twoside]{article}
\usepackage{fully3d}

\usepackage{subcaption}
\begin{document}

\title{A Machine-learning Based Initialization for Joint Statistical Iterative Dual-energy CT with Application to Proton Therapy} 

\author[1]{Tao Ge}
\author[1]{Maria Medrano}
\author[1]{Rui Liao}
\author[2]{David G. Politte}
\author[3]{Jeffrey F. Williamson}
\author[1]{Joseph A. O’Sullivan}

\affil[1]{Department of Electrical \& Systems Engineering,
          Washington University in St. Louis, St. Louis, United States}

\affil[2]{Mallinckrodt Institute of Radiology,
          Washington University in St. Louis, St. Louis, United States}

\affil[3]{Department of Radiation Oncology, Washington University in St. Louis, St. Louis, United States}

\maketitle
\thispagestyle{fancy}


\begin{customabstract}
Dual-energy CT (DECT) has been widely investigated to generate more informative and more accurate images in the past decades. For example, Dual-Energy Alternating Minimization (DEAM) algorithm achieves sub-percentage uncertainty in estimating proton stopping-power mappings from experimental 3-mm collimated phantom data. However, elapsed time of iterative DECT algorithms is not clinically acceptable, due to their low convergence rate and the tremendous geometry of modern helical CT scanners. A CNN-based initialization method is introduced to reduce the computational time of iterative DECT algorithms. DEAM is used as an example of iterative DECT algorithms in this work. The simulation results show that our method generates denoised images with greatly improved estimation accuracy for adipose, tonsils, and muscle tissue. Also, it reduces elapsed time by approximately 5-fold for DEAM to reach the same objective function value for both simulated and real data. 
\end{customabstract}


\section{Introduction}

Over the last 20 years, proton radiotherapy has been increasingly used to treat certain types of cancers because it has fewer side effects than traditional radiation. The absorbed dose reaches the maximum value after the proton beam traveling a certain depth in the object and then drops to near zero immediately. This peak is relatively narrow and is known as the Bragg peak. The estimated proton stopping power ratio (SPR) mapping allows radiation oncologists to align the Bragg peak to the tumor of the object, irradiating diseased tissues while sparing healthy cells. The current clinical practice estimates the SPR mappings from the single-energy CT (SECT) results, which leads to $2-3.5\%$ proton beam range uncertainty.

Dual-energy CT (DECT) SPR estimation methods were introduced to reduce the SPR uncertainty. Previous studies have shown that our iterative DECT algorithm, dual-energy alternating minimization (DEAM), has achieved sub-percentage uncertainty in estimating proton stopping-power mappings from experimental 3 mm collimated phantom data \cite{Maria20}.

However, iterative DECT algorithms are quite time-consuming when reconstructing 3D image volumes from helical sinograms, due to the large system operator and its low convergence rate. Compared to SECT, DECT algorithms reconstruct two measured sinograms scanned at different peak energies, which at least doubles the required number of system operations per iteration. Moreover, because the objective function of DEAM is decoupled in two more domains than the objective function in single-energy monoenergetic CT alternating minimization (AM), DEAM converges much slower than the AM algorithm with respect to the number of iterations. These factors make it difficult to get an accurate DECT result within a clinically acceptable time (20 minutes). 

\begin{figure}
  \centering
  \includegraphics[width=0.5\textwidth]{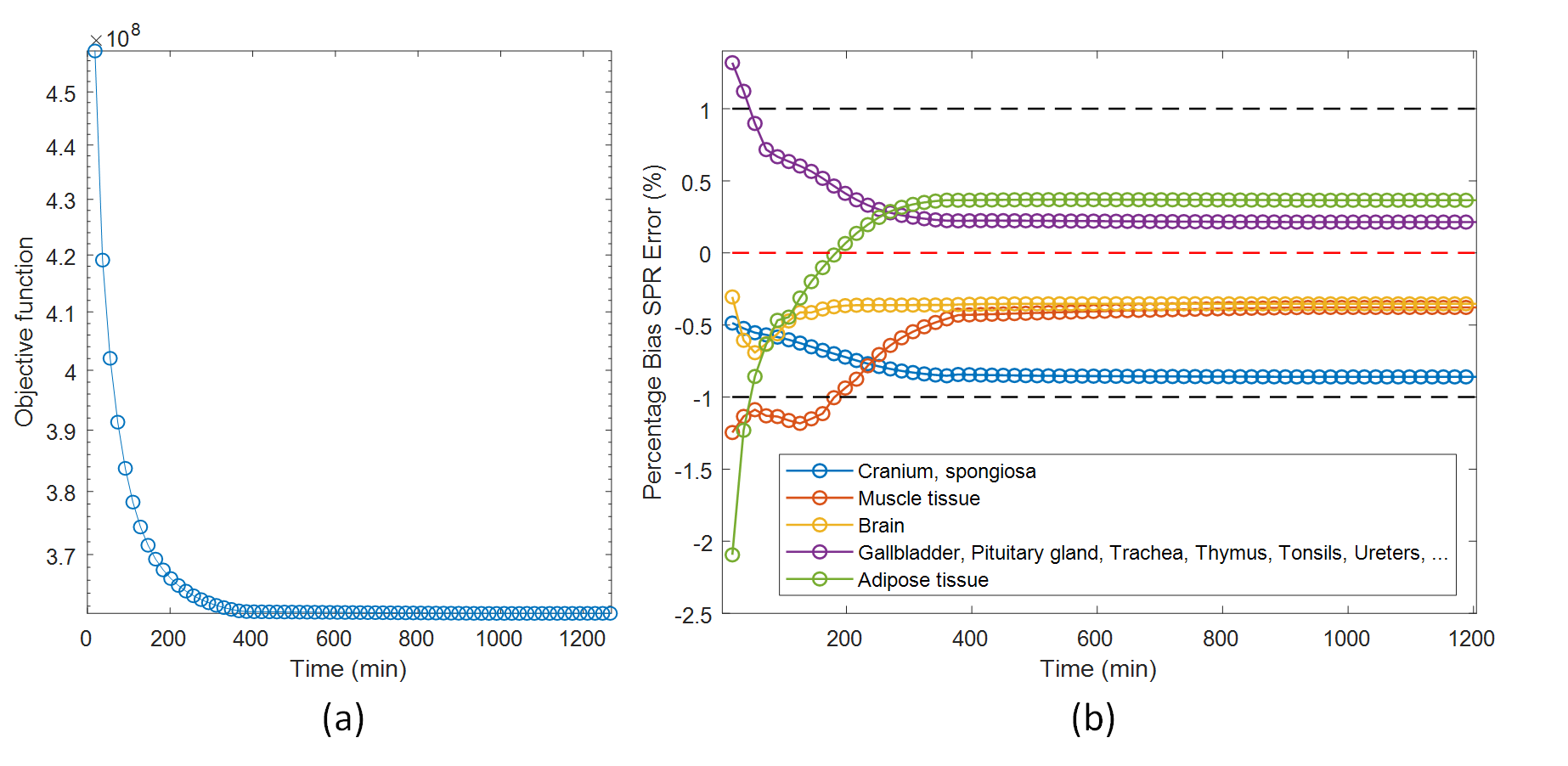}
  \caption{(a) plot objective function of DEAM versus time, (b) percentage biases of the SPR mapping derived from DEAM results with different elapsed time}
  \label{fig:spr_plot}
\end{figure}

Several algorithm-based and implementation-based acceleration methods have been taken into account, including GPU computation and the ordered subsets method, but it still takes more than 4 hours for the DEAM algorithm to converge. Figure \ref{fig:spr_plot}(a) shows the plotted objective function of DEAM versus time in minutes. It can be seen that the objective function converges after around 400 minutes. Figure \ref{fig:spr_plot}(b) shows the percentage biases of the SPR mapping derived from DEAM results with different elapsed time. The SPR biases are calculated inside 5 regions of interest, corresponding to 5 different materials in a virtual human head phantom. It can be seen that the percentage biases reach the $(-1\%,+1\%)$ range after around 180 minutes and become steady after around 380 minutes, $9\times$ or $19\times$ greater than the target elapsed time. 

Neural networks have been widely used in image processing and reconstruction because they are able to learn complicated potential image features which are difficult to capture with model-based methods. However, CNNs do not reliably generate accurate and critical information due to data susceptibility. As a result, they have been combined with the model-based optimization method to take advantage of the known physics knowledge. In \cite{Venkatakrishnan2013,Yu2018}, researchers plug a pre-trained denoising CNN as a prior into a model-based optimization algorithm to solve different inverse imaging problems. In \cite{Monakhova19}, an unrolled network of the model-based algorithm is constructed with trained hyper-parameters and a CNN regularizer for mask-based lensless imaging. In \cite{Hauptmann2018}, photoacoustic tomography images are updated iteratively by a pre-trained CNN based on the previous image volume and the gradient computed by the model-based algorithm. In this paper, We introduce a CNN-based initialization method to better estimate the initial condition of DEAM, which takes advantage of CNN's speedup while sparing the data susceptibility.

\section{Materials and Methods}

\subsection{Dual Energy Alternating Minimization (DEAM)}
DEAM is a joint statistical iterative algorithm that minimizes the objective function given by the sum of I-divergence \cite{o07}, 
\begin{equation}
I(d||g)=\sum_j d_j(y)\ln\frac{d_j(y)}{g_j(y:c)}-d_j(y)+g_j(y:c),
\end{equation}
and a penalty term, 
\begin{equation}
R(c)=\lambda \sum_{i=1}^2\sum_x\sum_{\tilde{x}\in N_x} w(x,\tilde{x}) \Phi\left(c_i(x)-c_i(\tilde{x})\right),
\end{equation}
\begin{equation}
\Phi(t) = \delta^2\left(\left|\frac{t}{\delta}\right|+\log\left(1-\left|\frac{t}{\delta}\right|\right)\right),
\end{equation}
where $x$, $y$ denote the indices of the discretized image space and measurement space, respectively. $N_x$ denotes the set of the neighbouring voxels of the image index $x$, $w(x,\tilde{x})$ is the voxel weight calculated as the inverse physical distance between voxel $x$ and $\tilde{x}$ $\lambda$ and $\delta$ are two hyper-parameters that control the weight and sparsity of the regularization term, $i$ denotes image component index (specifically $1$ for polystyrene and $2$ for CaCl2 ), $j$ denotes measured data index (specifically $1$ for 90 kVp and $2$ for 140 kVp ), $d$ denotes measured data, $g(y:c)$ denotes the estimation of measured data based on image components $c_i$, which is the forward model, written as
\begin{equation}
g_j(y:c)=\sum_E I_{0,j}(y,E)\exp\left(-\sum_x h(x,y)\sum_{i=1}^2 \mu_i(E)c_i(x)\right),
\end{equation}
where $\mu_i(E)$ denotes the attenuation coeffcient of the $i^{th}$ material at energy $E$, $I_{0,j}$ denotes the photon counts of the $j^{th}$ peak energy in the absence of an object, which contains information of the spectrum and the bowtie filter, and $h(x,y)$ denotes the system operator that represents the helical fan beam CT system.\\
The original initial condition was estimated by a iterative filtered backprojection (iFBP) based algorithm \cite{Chye2000} which requires less computational resources but has less in-practice accuracy than DEAM algorithm.

\begin{figure*}
  \centering
  \includegraphics[width=0.8\textwidth]{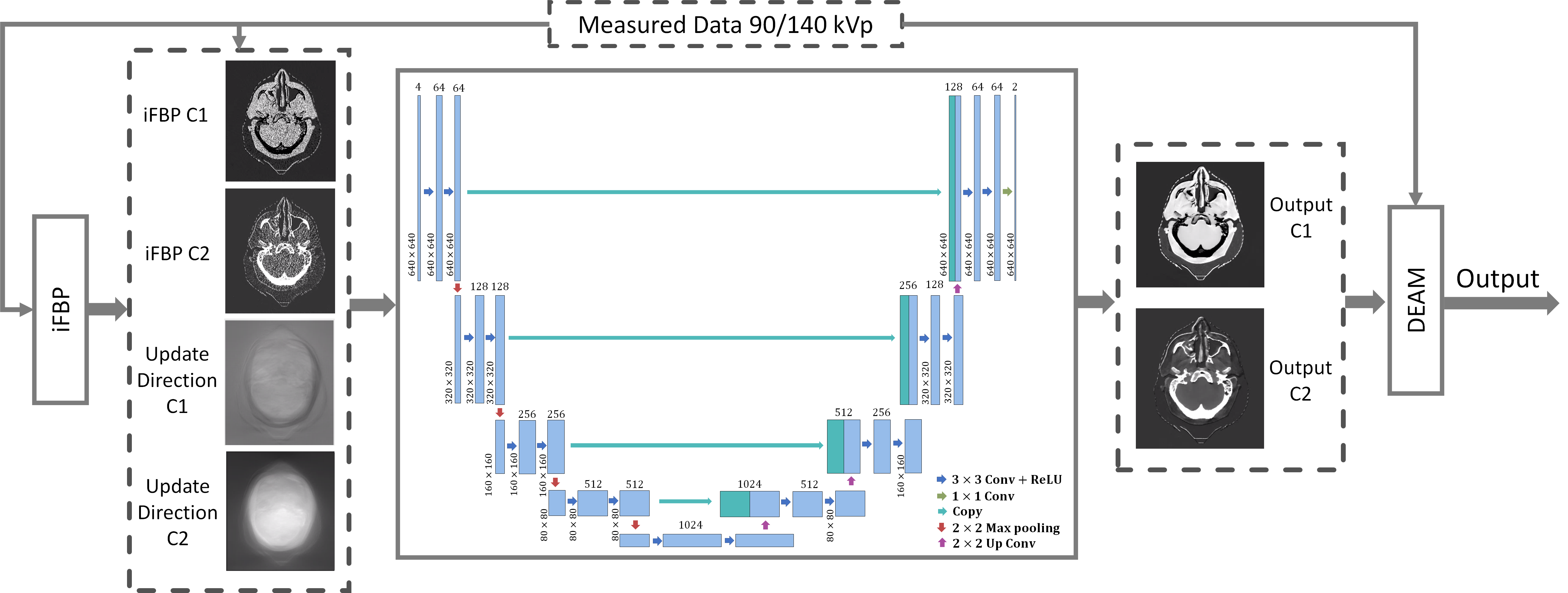}
  \caption{The overall process of DEAM initialization.}
  \label{fig:CNN}
\end{figure*}

\subsection{CNN-based initialization method}
The architecture of the whole CNN-based estimation process is shown in figure \ref{fig:CNN}. The main idea is to utilize CNN to estimate a better initial guess based on iFBP, our previous initialization method for DEAM. The proposed CNN has a widely used U-net structure originally from \cite{RonnebergerFB15} with some modifications. Its encoding-decoding structure allows the neural network to learn global features and reduce noise. This CNN takes four inputs: iFBP $c_1$ image, iFBP $c_2$ image, the DEAM update direction of $c_1$, and the DEAM update direction of $c_2$. It has been observed that the CNN trained with update directions performs better than the CNN trained without update directions. All the input and output images share the same positioning and sampling information, so the entire process could take advantage of the image alignment. Due to the computational cost and their physical property, each slice of the 3D reconstructed image is trained or tested separately. The reconstructed image with the size $610\times 610$ is zero-padded to $640\times 640$ pixels to fit this U-net structure. \\
The update directions for two basis vector model (BVM) components $c_1, c_2$ read
\begin{equation}
ud_i (x) = \log \frac{\sum_j p^B_{ij}(x)}{\sum_j q^B_{ij}(x)},
\end{equation}
where $p^B_{ij}$ denotes the backprojection of the $j^{th}$ measured data recalculated based on BVM and the spectrum for the $i^{th}$ basis, and $q^B_{ij}$ denotes the backprojection of the $j^{th}$ estimation data recalculated based on BVM and the spectrum for the $i^{th}$ basis. Therefore, the update direction gives the pixel-wise distance of the estimated components to the ``truth.'' \\
In unregularized DEAM, two basis vector model components $c_1, c_2$ are updated by 
\begin{equation}
c^{k+1}_i(x) = c^{k}_i(x) - \frac{1}{Z_i(x)}\log \frac{\sum_j p^{B,k}_{ij}(x)}{\sum_j q^{B,k}_{ij}(x)},
\end{equation}
where $Z_i(x)$ is the auxiliary variable to ensure convergence. \\
Then, the CNN estimation
\begin{equation}
c_1,c_2 = CNN(c_1,c_2,ud_1,ud_2)
\end{equation}
could be regarded as an update step of DEAM whose step size and regularization term are trained rather than embedded. 

\section{Results and Discussion}
\subsection{Simulation results}
\begin{figure}
  \centering
  \includegraphics[width=0.5\textwidth]{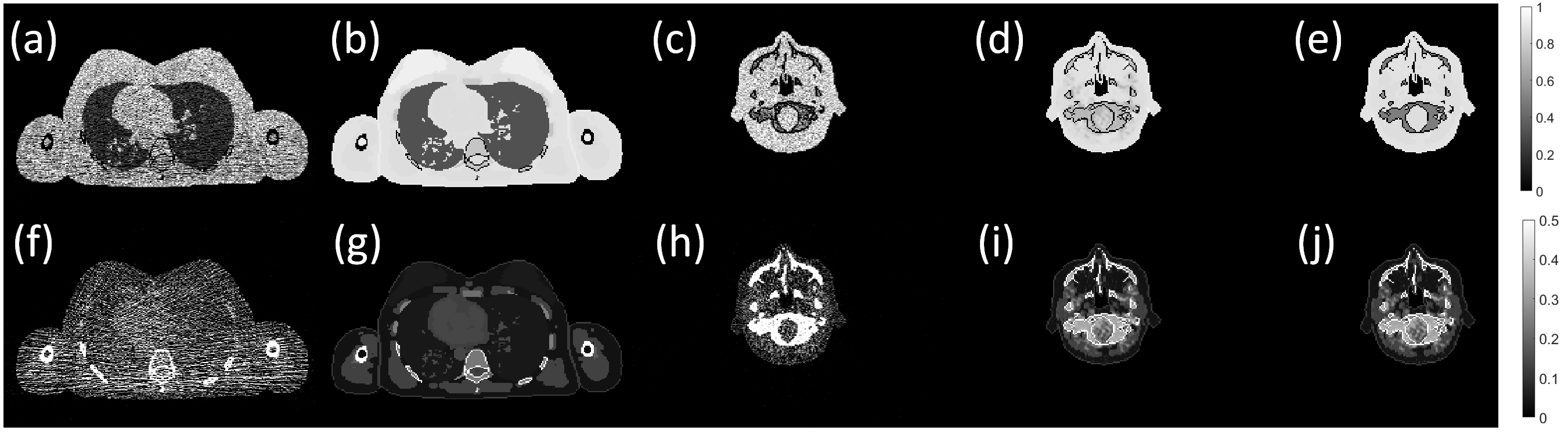}
  \caption{Examples of simulation images. From left column to right column: training input, training output, test input, test output, test ground truth (as reference). Top row: C1; Bottom row: C2.}
  \label{fig:cnn_img}
\end{figure}

\begin{figure}
  \centering
  \includegraphics[width=0.5\textwidth]{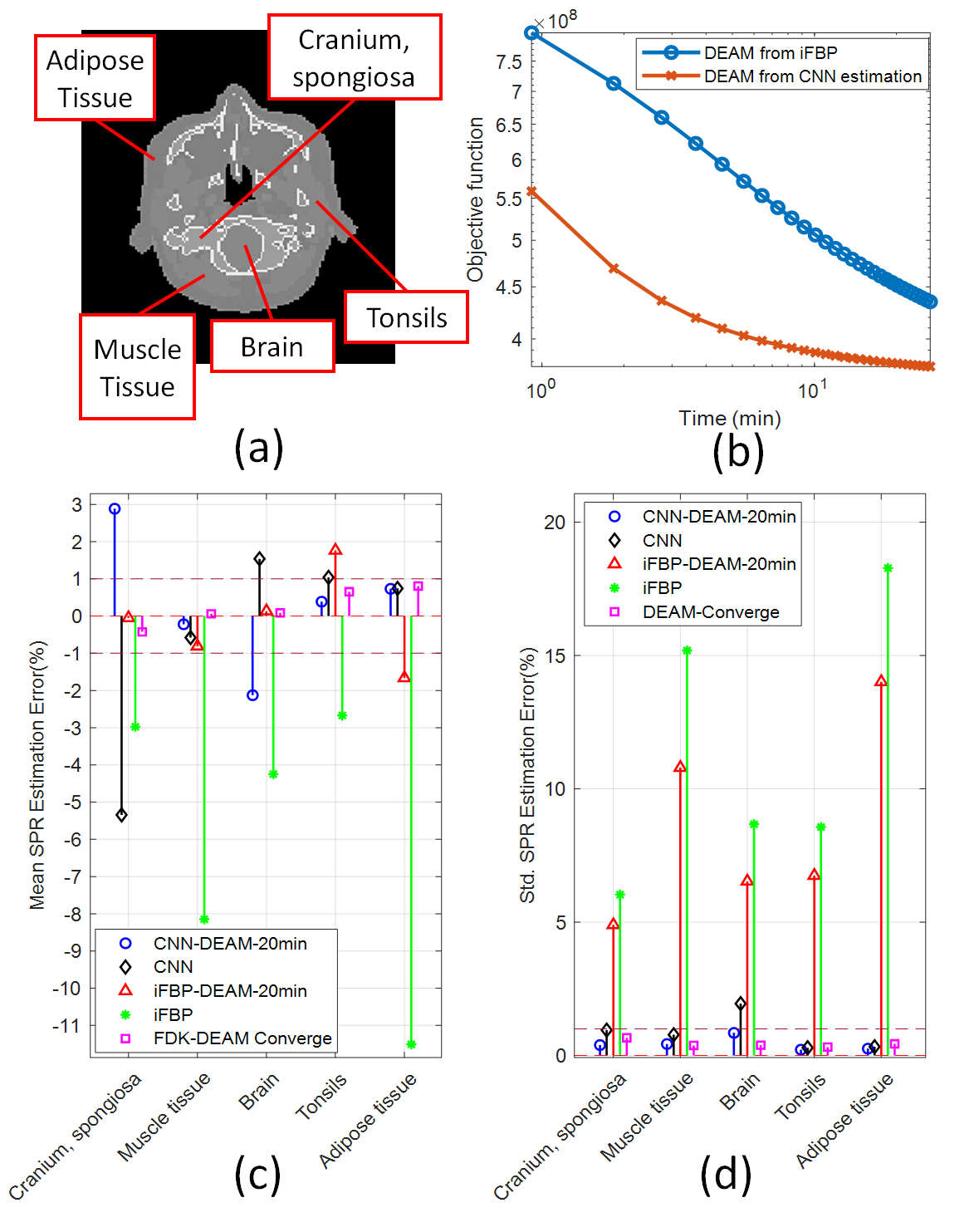}
  \caption{(a) Five ROIs in different materials, (b) plot of objective function of DEAM starting from iFBP result and DEAM starting from CNN estimation vs. time, (c) and (d) mean SPR estimation percentage error and SPR standard deviation of DEAM 20 minutes result with CNN estimation as initial condition, iFBP result, DEAM 20 minutes result with iFBP estimation as initial condition and DEAM converged result.}
  \label{fig:spr_cnn}
\end{figure}

The measured data is simulated from the ICRP \cite{ICRP}, a virtual female phantom. The simulation system operator has the same geometry as the Phillips Brilliance Big Bore CT scanner. The reconstruction and CNN training-testing process are done in a 20-threaded computer with 4 GTX 1080TI. The training process takes the iFBP result of ICRP chest and its corresponding update directions as the input and takes the ground truth of ICRP chest as the output. In the test process, the input of the trained CNN is the iFBP result of ICRP head. Figure \ref{fig:cnn_img} shows an example slice of the training and testing data. Chest iFBP is noisier than head iFBP since the larger area of the chest leads to the larger line integral of attenuation coefficients. Compared the CNN head result to the ground truth, the CNN eliminates most of the noise of iFBP images but introduces some artifacts.\\
Figure \ref{fig:spr_cnn} shows the quantitative comparison. Five regions of interest are selected for different materials and indicated by red lines in \ref{fig:spr_cnn}(a). \ref{fig:spr_cnn}(b) shows the objective function of DEAM starting from different initial condition versus time. The objective function of iFBP is initially around 10 times larger than the objective function of CNN estimation. The objective function of CNN estimation after 3 minutes is close to the objective function of iFBP after 25 minutes, which means CNN introduces $\sim8\times$ speedup in this case. \ref{fig:spr_cnn}(c) and (d) are the plots of percentage bias and percentage standard deviation of SPR mapping estimated by 20 minutes DEAM result starting from CNN estimation, the result of CNN estimation, iFBP, 20 minutes DEAM result starting from iFBP, the result of iFBP and converged DEAM. \\
Converged DEAM has the best overall performance. The uncertainty measures of converged DEAM for all ROIs are within 1\%. iFBP has the worst overall performance with the largest uncertainty measures. CNN and CNN-DEAM-20min have less bias than iFBP-DEAM-20min in muscule tissue, tonsils, and adipose tissue, but the bias of CNN and CNN-DEAM-20min beyond $\pm 1\%$ range in spongiosa (cranium) and brain. Due to the absence of brain tissue and spongiosa (cranium) in training data, it is not surprising that CNN fails to estimate their BVM components. We hypothesize that it could be fixed by containing brain slices in the training data.   \\
iFBP got the highest standard deviation, which matches the noise level of iFBP images. Standard deviation of iFBP-DEAM-20min is the second highest, because the data fidelity term is much greater than the penalty term in early iterations, so DEAM tends to reduce the data fidelity term rather than the penalty term. CNN and CNN-DEAM-20min did a great job in the standard deviation analysis since the U-net structure is good at denoising. 
\subsection{Clinic results}
\begin{figure}
  \centering
  \includegraphics[width=0.5\textwidth]{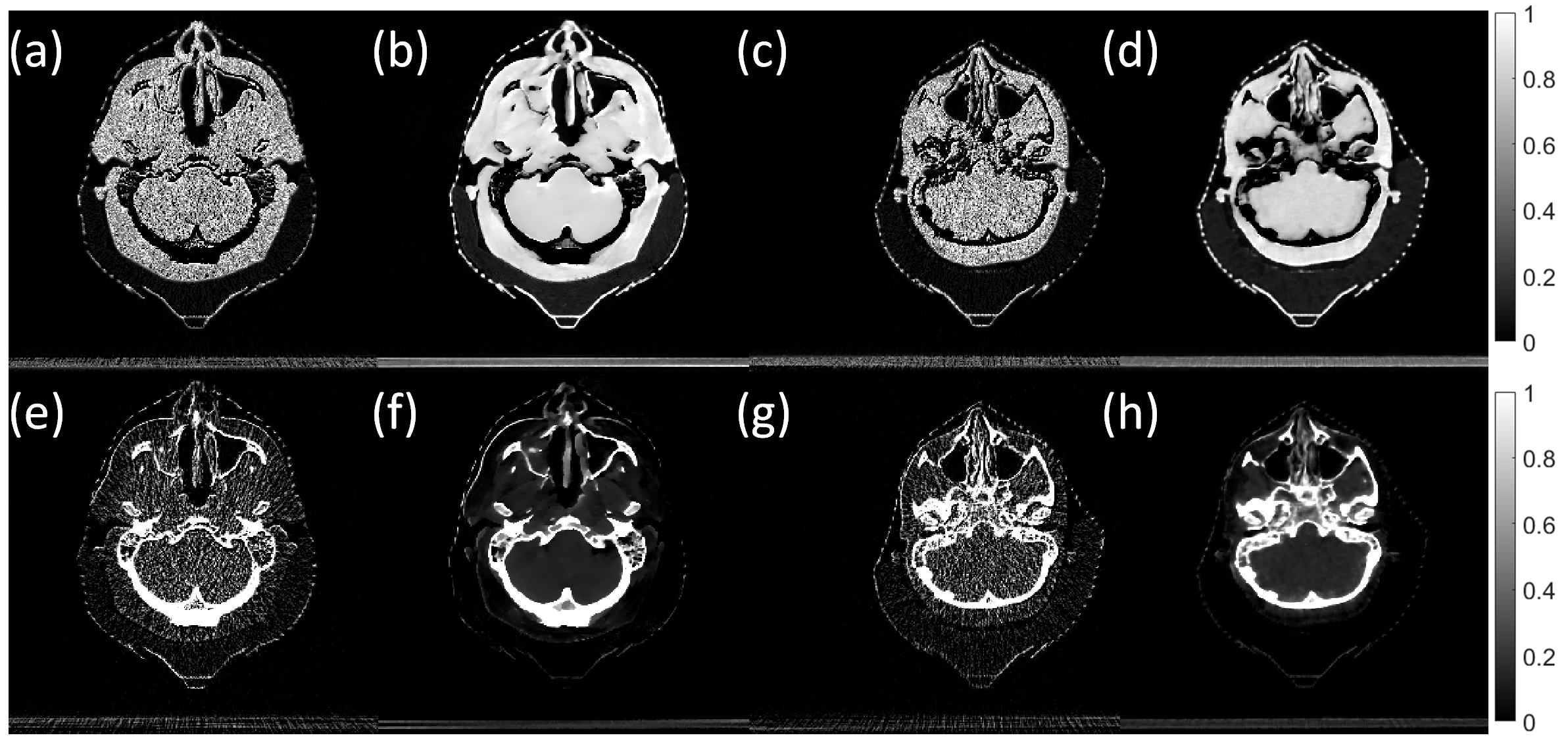}
  \caption{Examples of real patient images. From left column to right column: training input, training output, test input, test output. Top row: C1; Bottom row: C2}.
  \label{fig:cnn_img_real}
\end{figure}

\begin{figure}
  \centering
  \includegraphics[width=0.5\textwidth]{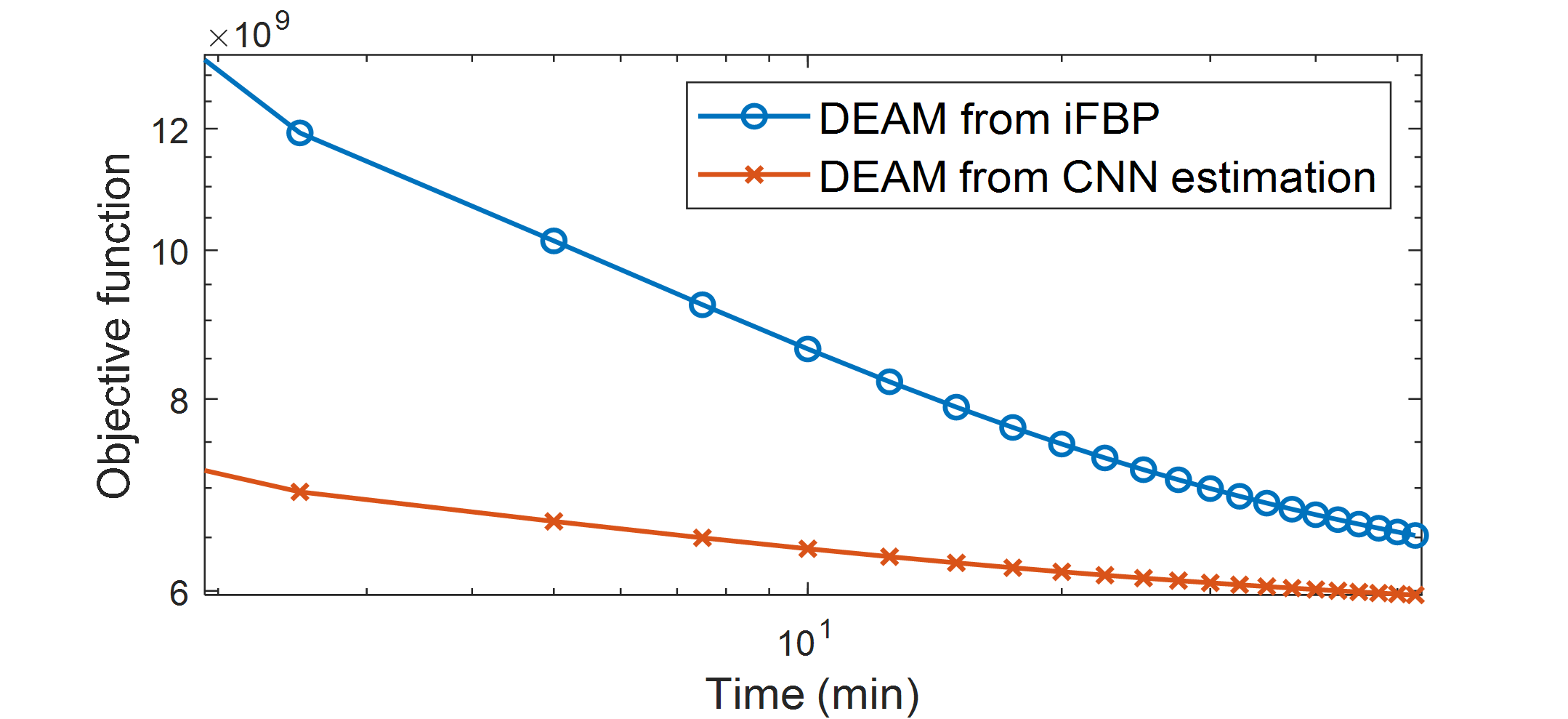}
  \caption{Real patient objective function of DEAM starting from iFDK and CNN estimation vs. time.}
  \label{fig:obj_cnn_real}
\end{figure}

In this section, all 90 kVp and 140 kVp clinically measured data are acquired sequentially on the Phillips Brilliance Big Bore CT scanner, with 12 mm collimation. The CNN takes iFBP of the head scan as the training input and DEAM converged result of the same head scan as the training input. In the test process, iFBP of the head scan of a different patient is used as the input. The example images of the experimental training and testing data are shown in figure \ref{fig:cnn_img_real}.\\
Figure \ref{fig:obj_cnn_real} shows the plot of the real patient objective function of DEAM starting from iFDK and CNN estimation versus time. Similar to simulation results, the objective function of CNN estimation after 8 minutes is close to the objective function of iFBP after 70 minutes, which means CNN also introduces $\sim8\times$ speedup in the real patient case. 

\section{Conclusion}
We have proposed a CNN-based method that improves the initial guess for DEAM. We only show the CNN-based initialization of one iterative algorithm, but it is applicable for other iterative algorithms. This CNN initialization method takes iFBP result as the input, generating a denoised image with a great improvement of estimation uncertainty for adipose, tonsils, and muscle tissue in the simulation task. However, the method did not work well in estimating brain and spongiosa tissue, due to the absence of brain and spongiosa (cranium) in training data. We hypothesize that the misestimation of spongiosa and brain tissse could be fixed by containing brain slices in the training data. It is desirable to use CNN trained with head-neck images for head-neck data, and CNN trained with thorax images for thorax data. Also, in both the simulation and real patient tasks, the proposed method reduces elapsed time approximately 8-fold for DEAM to reach the same objective function value.

\section{Acknowledgments}
This study is supported by NIH R01 CA 212638.

\printbibliography

@inproceedings{Maria20,
author = {Maria Medrano and Ruirui Liu and Tyler Webb and Tianyu Zhao and Jeffrey Williamson and Bruce Whiting and David G. Politte and Mariela Porras-Chaverri and Joseph A. O'Sullivan},
title = {{Accurate proton stopping power images reconstructed using joint statistical dual energy CT: experimental verification and impact of fan-beam CT scatter}},
volume = {11312},
booktitle = {Medical Imaging 2020: Physics of Medical Imaging},
editor = {Guang-Hong Chen and Hilde Bosmans},
organization = {International Society for Optics and Photonics},
publisher = {SPIE},
pages = {430 -- 438},
year = {2020},
doi = {10.1117/12.2549788},
URL = {https://doi.org/10.1117/12.2549788}
}

@INPROCEEDINGS{ICRP,
  author={ICRP},
  booktitle={ICRP Publication 110. Ann. ICRP. }, 
  title={Adult Reference Computational Phantoms}, 
  year={2009},
  volume={39(2)}}

@ARTICLE{o07,
  author={J. A. {O'Sullivan} and J. {Benac}},
  journal={IEEE Transactions on Medical Imaging}, 
  title={Alternating Minimization Algorithms for Transmission Tomography}, 
  year={2007},
  volume={26},
  number={3},
  pages={283-297},
  doi={10.1109/TMI.2006.886806}}

@ARTICLE{Hauptmann2018,
  author={A. {Hauptmann} and F. {Lucka} and M. {Betcke} and N. {Huynh} and J. {Adler} and B. {Cox} and P. {Beard} and S. {Ourselin} and S. {Arridge}},
  journal={IEEE Transactions on Medical Imaging}, 
  title={Model-Based Learning for Accelerated, Limited-View 3-D Photoacoustic Tomography}, 
  year={2018},
  volume={37},
  number={6},
  pages={1382-1393},
  doi={10.1109/TMI.2018.2820382}}

@INPROCEEDINGS{Venkatakrishnan2013,
  author={S. V. {Venkatakrishnan} and C. A. {Bouman} and B. {Wohlberg}},
  booktitle={2013 IEEE Global Conference on Signal and Information Processing}, 
  title={Plug-and-Play priors for model based reconstruction}, 
  year={2013},
  volume={},
  number={},
  pages={945-948},
  doi={10.1109/GlobalSIP.2013.6737048}}

@article{Yu2018,
  author    = {Yu Sun and
               Brendt Wohlberg and
               Ulugbek S. Kamilov},
  title     = {An Online Plug-and-Play Algorithm for Regularized Image Reconstruction},
  journal   = {CoRR},
  volume    = {abs/1809.04693},
  year      = {2018},
  url       = {http://arxiv.org/abs/1809.04693},
  archivePrefix = {arXiv},
  eprint    = {1809.04693},
  bibsource = {dblp computer science bibliography, https://dblp.org}
}

@article{Monakhova19,
author = {Kristina Monakhova and Joshua Yurtsever and Grace Kuo and Nick Antipa and Kyrollos Yanny and Laura Waller},
journal = {Opt. Express},
number = {20},
pages = {28075--28090},
publisher = {OSA},
title = {Learned reconstructions for practical mask-based lensless imaging},
volume = {27},
month = {Sep},
year = {2019},
url = {http://www.opticsexpress.org/abstract.cfm?URI=oe-27-20-28075},
doi = {10.1364/OE.27.028075}
}

@article{RonnebergerFB15,
  author    = {Olaf Ronneberger and
               Philipp Fischer and
               Thomas Brox},
  title     = {U-Net: Convolutional Networks for Biomedical Image Segmentation},
  journal   = {CoRR},
  volume    = {abs/1505.04597},
  year      = {2015},
  url       = {http://arxiv.org/abs/1505.04597},
  archivePrefix = {arXiv},
  eprint    = {1505.04597},
  bibsource = {dblp computer science bibliography, https://dblp.org}
}

@article{Chye2000,
  author={ {Chye Hwang Yan} and R. T. {Whalen} and G. S. {Beaupre} and S. Y. {Yen} and S. {Napel}},
  journal={IEEE Transactions on Medical Imaging}, 
  title={Reconstruction algorithm for polychromatic CT imaging: application to beam hardening correction}, 
  year={2000},
  volume={19},
  number={1},
  pages={1-11},
  doi={10.1109/42.832955}}

\end{document}